\documentclass[twoside]{article}
\usepackage{graphicx}
\usepackage{hyperref}
\usepackage{natbib}

\usepackage[accepted]{aistats2021}
\begin{document}

%

%

\twocolumn[

\aistatstitle{CLAR: Contrastive Learning of Auditory Representations}

\aistatsauthor{ Haider Al-Tahan \And Yalda Mohsenzadeh}

\aistatsaddress{ Department of Computer Science \\ The University of Western Ontario \\ London, Ontario \\ Canada \And  Department of Computer Science \\ The University of Western Ontario \\ London, Ontario \\ Canada } ]

\begin{abstract}
Learning rich visual representations using contrastive self-supervised learning has been extremely successful. However, it is still a major question whether we could use a similar approach to learn superior auditory representations. In this paper, we expand on prior work (SimCLR) to learn better auditory representations. We (1) introduce various data augmentations suitable for auditory data and evaluate their impact on predictive performance, (2) show that training with time-frequency audio features substantially improves the quality of the learned representations compared to raw signals, and (3) demonstrate that training with both supervised and contrastive losses simultaneously improves the learned representations compared to self-supervised pre-training followed by supervised fine-tuning. We illustrate that by combining all these methods and with substantially less labeled data, our framework (CLAR) achieves significant improvement on prediction performance compared to supervised approach. Moreover, compared to self-supervised approach, our framework converges faster with significantly better representations. 
\end{abstract}

\section{INTRODUCTION}

Although humans are proficient at perceiving and understanding sounds, making algorithms perform the same task poses a challenge due to the wide range of variations in auditory features. Applications of sound understanding range from surveillance \citep{radhakrishnan2005audio} and music classification \citep{choi2017convolutional, 9054352} to audio generation \citep{engel2018gansynth, donahue2019adversarial} and deep-fake detection \citep{mittal2020emotions}.

Achieving automated auditory perception requires the learning of effective representations. Often prior work derive effective representations through discriminative approaches \citep{park2019specaugment, hershey2017cnn, tokozume2017learning, Guzhov2020ESResNetES}. That is, similar to supervised learning, the model learns the mapping between the input signal to the class label. The underlying assumption with such approach is that the latent representations carry effective representations for the designed tasks. One fundamental problem with such learned representations is the potential limitation to generalizability. First, those representations are only limited to availability of expensive and time consuming labeled data. Secondly, representations are skewed towards one particular domain (e.g. speech, music, etc \ldots). Therefore, in both cases, major fine-tuning to the targeted training data would be required. Alternatively, recent self-supervised approaches using contrastive learning in the latent space have been shown to learn efficient representations that achieves state-of-the-art performance in images \citep{chen2020simple, qian2020spatiotemporal, bachman2019learning, oord2018representation, dosovitskiy2014discriminative, hadsell2006dimensionality} and videos \citep{qian2020spatiotemporal}. However, it is still a major question on how we can achieve similar landmark on auditory data. 

In this work, we build on SimCLR \citep{chen2020simple}, a self-supervised framework for contrastive learning of visual representations. We show that similar framework could be adopted for learning effective auditory representations. Moreover, with a simple modification, we are able to reduce the training time and improve recognition performance.

In order to accomplish this, we introduce three major components that are important to nourish the learning of auditory representations. We:

\begin{itemize}
    \item Demonstrate the success of contrastive learning in learning efficient auditory representations.
    \item Investigate six data augmentation operations and show their effect on auditory classification task both with raw audio and extracted time-frequency audio features.
    \item Show that training with time-frequency audio features substantially improves the quality of the learned representations in contrastive learning compared to raw audio signals.
    \item Show that using supervised and contrastive learning simultaneously while training, not only improves the learned representations but also speeds up the training.
\end{itemize}

By combining all these methods, our framework (CLAR) achieves 96.1\% top-1 accuracy, which is 1\% relative improvement over supervised method and 11.3\% improvement over SimCLR on Speech Command dataset. Moreover, compared to SimCLR, our framework converges faster with significantly better representations. 

\section{RELATED WORK}

\subsection{Contrastive Learning}

In an era of ever increasing unlabeled data, contrastive learning has been shown to be effective at capitalizing on such data \citep{1640964, dosovitskiy2014discriminative, wu2018unsupervised, zhuang2019local}. Contrastive learning is a self-supervised framework that formulates representations in a given model based on similarity/dissimilarity of a given input pairs. Recently proposed method called SimCLR has been shown to not only outperform previous self-supervised methods on ImageNet but also outperform supervised methods on some natural image classification datasets \citep{chen2020simple}. In essence, the framework aims at learning efficient representations by maximizing agreement between differently augmented views of the same data and maximizing difference across contrasting images via contrastive loss in the latent space. The loss for such objective is termed Normalized Temperature-scaled Cross Entropy Loss (NT-Xent):

\begin{equation}
    l_{i,j} = -\log\frac{\exp\left(\text{sim}\left(\mathbf{z}_{i}, \mathbf{z}_{j}\right)/\tau\right)}{\sum^{2N}_{k=1}\mathbf{1}_{[k\neq{i}]}\exp\left(\text{sim}\left(\mathbf{z}_{i}, \mathbf{z}_{k}\right)/\tau\right)}
\end{equation}

where $\mathbf{1}_{[k\neq{i}]} \in 0, 1$ is an indicator function evaluating to $1$ iff $k\neq{i}$ and $\tau$ denotes a temperature parameter (default is 0.5). The loss is computed across all positive pairs, both $\left(i, j\right)$ and $\left(j, i\right)$, in a mini-batch. $\text{sim}\left(\mathbf{u}, \mathbf{v}\right) = \mathbf{u}^{T}\mathbf{v}/||\mathbf{u}|| ||\mathbf{v}||$ denote the cosine similarity between two vectors $\mathbf{u}$ and $\mathbf{v}$. 

Our work extends SimCLR by introducing an approach to learning auditory representations instead of visual representations. We do that by showing composition of augmentations that are most efficient in learning auditory representations. Previous work on auditory data augmentations have shown effective methods that improve supervised classifications applied both on raw audio signal \citep{mcfee2015software, schluter2015exploring, ko2015audio} and time-frequency audio features \citep{Park_2019}. However, auditory data augmentations that nourish effective representations with contrastive learning is yet to be investigated. In this work, we also investigate the effect of the augmentations on learning representations from both raw audio signal and time-frequency audio features in contrastive learning framework.

\subsection{Catastrophic Forgetting}

A common practice in deep learning is to fine-tune parameters of a model - sometimes trained on a completely different task - to improve performance. In a notable example, \cite{Guzhov2020ESResNetES} demonstrate that the initialization of weights based on pre-training on the ImageNet ILSVRC-2012 \citep{krizhevsky2012imagenet} image classification task improves performance for environmental sound classification. In self-supervised learning this approach is adopted by performing self-supervised pre-training followed by supervised fine-tuning on labeled examples \citep{chen2020simple, hnaff2019dataefficient, He_2020, kiros2015skipthought}. Although fine-tuning can be beneficial in big models (i.e. $>$ 100 Million Parameters), fine-tuning on smaller models may result in \textit{catastrophic forgetting} \citep{mccloskey1989catastrophic, goodfellow2013empirical, li2017learning}. Catastrophic forgetting is a situation where in learning new tasks, the model may not use the shared parameters from the old task and "forget" them in the process. This phenomenon could partially explain why contrastive learning benefits more from bigger models than its supervised counterpart \citep{chen2020simple}. That is, bigger models have the capacity to maintain representations from both pre and post fine-tuning. \cite{khosla2020supervised} demonstrate a promising approach to resolve this issue by modifying contrastive loss to leverage labeled information in learning efficient visual representation. However, such approach makes the contrastive loss strictly supervised and limit the method to labeled data. Our approach incorporate both supervised and self-supervised frameworks during training without fine-tuning which not only foster more efficient representations but also speed the training process. Furthermore, unlike \cite{khosla2020supervised}, our proposed method can leverage labeled and unlabeled data simultaneously during training. 
\section{METHODS}
\subsection{Audio Pre-porcessing}
\begin{figure}[h]
\begin{center}
\includegraphics[width=6cm,height=7cm]{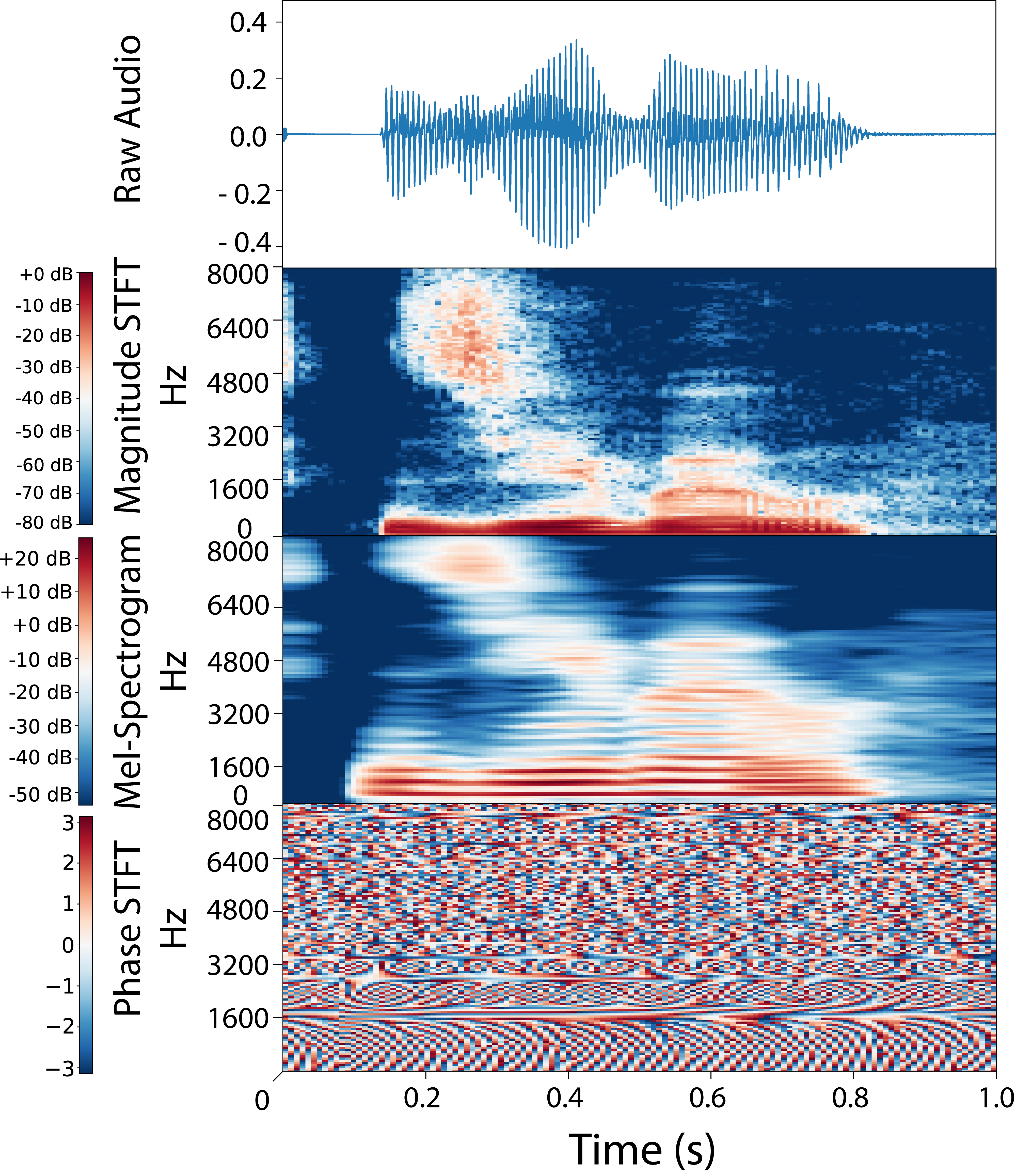}
\end{center}
\caption{Raw audio (first) with the subsequent time-frequency features, specifically, STFT (short-time Fourier transform) magnitudes (second) and phase angles (fourth), and Mel-spectorgram (third). The raw audio and time-frequency features were utilized in training 1D and 2D versions of ResNet. }\label{fig:demo}
\end{figure}

We trained two family of models, one that takes as input raw audio signals while the other utilizes time-frequency audio features as input. For both models, we start by down-sampling (when applicable) all audio signals to 16kHz, followed by signal padding (by zeros) or clipping the right side of the signal to ensure that all audio signals are of the same length. The target length of the audio signal is set based on the datasets' assigned audio length (Section \ref{subsec:datasets}). For models dependant on time-frequency features, we computed the short-time Fourier transform (STFT) magnitudes and phase angles of the input audio with 16 ms windows and 8 ms stride \citep{allen1977short}, further we projected the STFT to 128 frequency bins equally spaced on the Mel scale (Figure ~\ref{fig:demo}). Moreover, we computed the log-power of magnitude STFT and Mel spectrogram (Eq. \ref{logpower}) and stacked the three time-frequency features in the channel dimension, resulting in a matrix of size: $ 3 \times F \times T$. Where $F$ is the  number of the frequency bins and $T$ is the number of frames in the spectrogram. This was done to ensure that we have comprehensive features that could capture multi-domain audio signals (e.g. speech, environmental and music sounds) and would not require us to change the baseline ResNet 2D architecture. All spectrogram transformations were performed on GPU using 1D Convolutional Neural Networks \citep{cheuk2020nnaudio}.

\begin{equation}
    \label{logpower}
    f(S) = 10 log_{10}|S|^{2}
\end{equation}

where $S$ is the mel-spectrogram or magnitude STFT.

\subsection{Training/Evaluation Protocol}\label{sec:train}

To maintain consistency and allow for a fair comparison across supervised, self-supervised and our proposed method (CLAR), we adopted three components in our framework (see Figure \ref{fig:methods}): 

\begin{enumerate}
    \item \textbf{Encoder} extracts representations from data samples. To vectorize our representations we performed average pooling on the output of the encoder. For our encoder, we trained 1D and 2D variants of standard ResNet18 \citep{he2016deep} with SimCLR training protocol. For models trained on time-frequency audio features (i.e. spectrograms) as input, we utilized a typical ResNet18 \citep{hershey2017cnn} with random initialization. Alternatively, we switched all ResNet18 operations such as convolutions, max-pooling and batch normalization from 2D to 1D for models that takes raw audio signal as input. Furthermore, we squared the kernel size of 1D operations, to match the same number of parameters as 2D variant of the model.
    \item \textbf{Projection head} maps the extracted encoder representations to a space where contrastive loss is applied. Projection head consists of three fully connected layers with ReLU activation functions. We performed contrastive loss on the output of the projection head with fixed vector size of 128. In the supervised approach we replaced the vector used for contrastive loss with a vector with the same size as the number of classes to apply cross entropy loss. Lastly, in our proposed approach, in addition to the contrastive loss, we included an additional layer that maps the 128 vector to the number of classes where the cross entropy loss was applied.
    \item \textbf{Evaluation head} is similar to the projection head in terms of architecture but it was trained on top of the frozen encoder to assess the learned representation quality by computing the test accuracy. When limiting labeled data to compare performance across supervised, self-supervised and our proposed approach (Section \ref{sec4}), we trained the evaluation head on the full labeled data. This approach is commonly adopted to evaluate the learned representations of self-supervised methods \citep{chen2020simple, kolesnikov2019revisiting, bachman2019learning}. 
\end{enumerate}

We trained all models with 1024 batch size, layer-wise adaptive rate Scaling (LARS) optimizer \citep{you2017large} with learning rate of 1.0, weight decay of $10^{-4}$, linear warmup for the first 10 epochs, decay of the learning rate with the cosine decay schedule without restarts \citep{loshchilov2016sgdr} and global batch normalization. For some datasets in sections \ref{sec3}, we reduced the batch size to 512 to be able to fit the data in memory. In section \ref{sec4} we trained all our models on augmentations that we found to yield the best performance on test accuracy (section \ref{sec2} \& \ref{sec3}). All models were trained from random initialization with 4 NVIDIA v100 Tesla 32GB GPUs. 

\subsection{CLAR Framework}\label{sec:clar}

\begin{figure*}[h]
\begin{center}
\includegraphics[width=\linewidth]{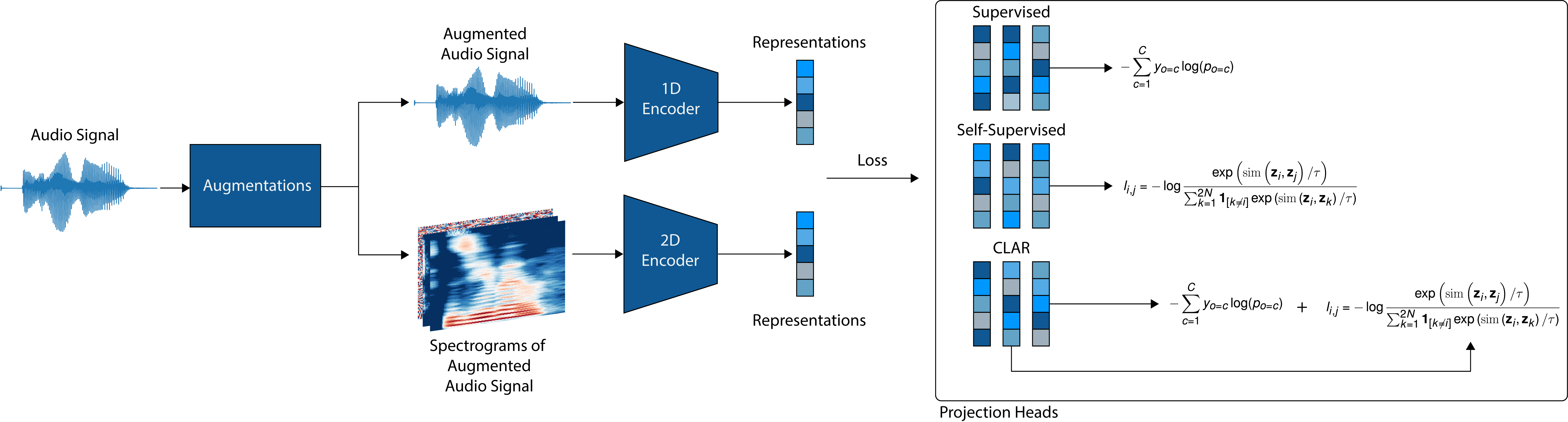}
\end{center}
\caption{Training pipeline. We start by applying augmentations directly to the audio signal followed by either feeding the augmented audio signal or the spectrograms of the augmented audio signal to the subsequent encoder. We feed the representations from the encoder to the same projection head architecture across all approaches. However, we change the loss and the the layer representation employed in the loss computation.}\label{fig:methods}
\end{figure*}

Contrastive and supervised learning share the common goal of constructing representations that distinguish samples for different tasks. The supervised approach focuses on distinguishing samples from multiple classes without constraints on the latent representations. While, contrastive learning constructs such representations between pairs of samples with the constraint being that latent representations of negative samples are maximized and positive samples are minimized. These two frameworks have their own advantages and disadvantages. For instance, constrastive learning benefits from larger batch sizes and longer training \citep{goyal2018accurate, chen2020simple}. However, the supervised approach is simpler to optimize, hence, requires less training to achieve relative performance \citep{chen2020simple}. With contrastive learning, the aim is to benefit from both frameworks by combining the shared representations from both frameworks by performing self-supervised pre-training followed by supervised fine-tuning on labeled examples \citep{chen2020simple, hnaff2019dataefficient, He_2020, kiros2015skipthought}. However, as allotted earlier this could result in \textit{catastrophic forgetting}, especially in smaller networks \citep{mccloskey1989catastrophic, goodfellow2013empirical, li2017learning}. In this work, we abolish the fine-tuning step and integrate both contrastive and supervised learning frameworks simultaneously during training:

\begin{equation}
    L =  \mbox{NT-Xent } - \sum_{i}^Ct_{i,o}\log(p_{o})
\end{equation}

where the first term (NT-Xent) is the contrastive loss and the second term is the Categorical Cross-Entropy (CE) loss of the labeled samples. In CE loss, $C$ is the number of classes, $t_i$ is a binary indicator (i.e. 0 or 1) if observed class label $o$ is the same as the class label $i$ and $p$ is the predicted probability of the observation $o$. In cases where the labels for some of the samples within the mini-batch are missing, then the CE loss will be set to zero. Alternatively, contrastive loss is always applied as it is not dependent on the labels. A possible approach to guarantee labeled samples within a mini-batch is to use stratified sampling, this is especially important when the labeled data is substantially small portion of the whole dataset (e.g. 1\% of the data is labeled). In our analysis, we utilized random sampling because (1) datasets utilized are not large, especially when compared with ImageNet and (2) we utilized very large batch size (1024). Using the projection head, we apply the CE loss on the last layer and the contrastive loss on the layer preceding the last layer (Figure \ref{fig:methods}).

\subsection{Augmentations}\label{sec:aug}

\begin{figure}[h]
\begin{center}
\includegraphics[width=\linewidth]{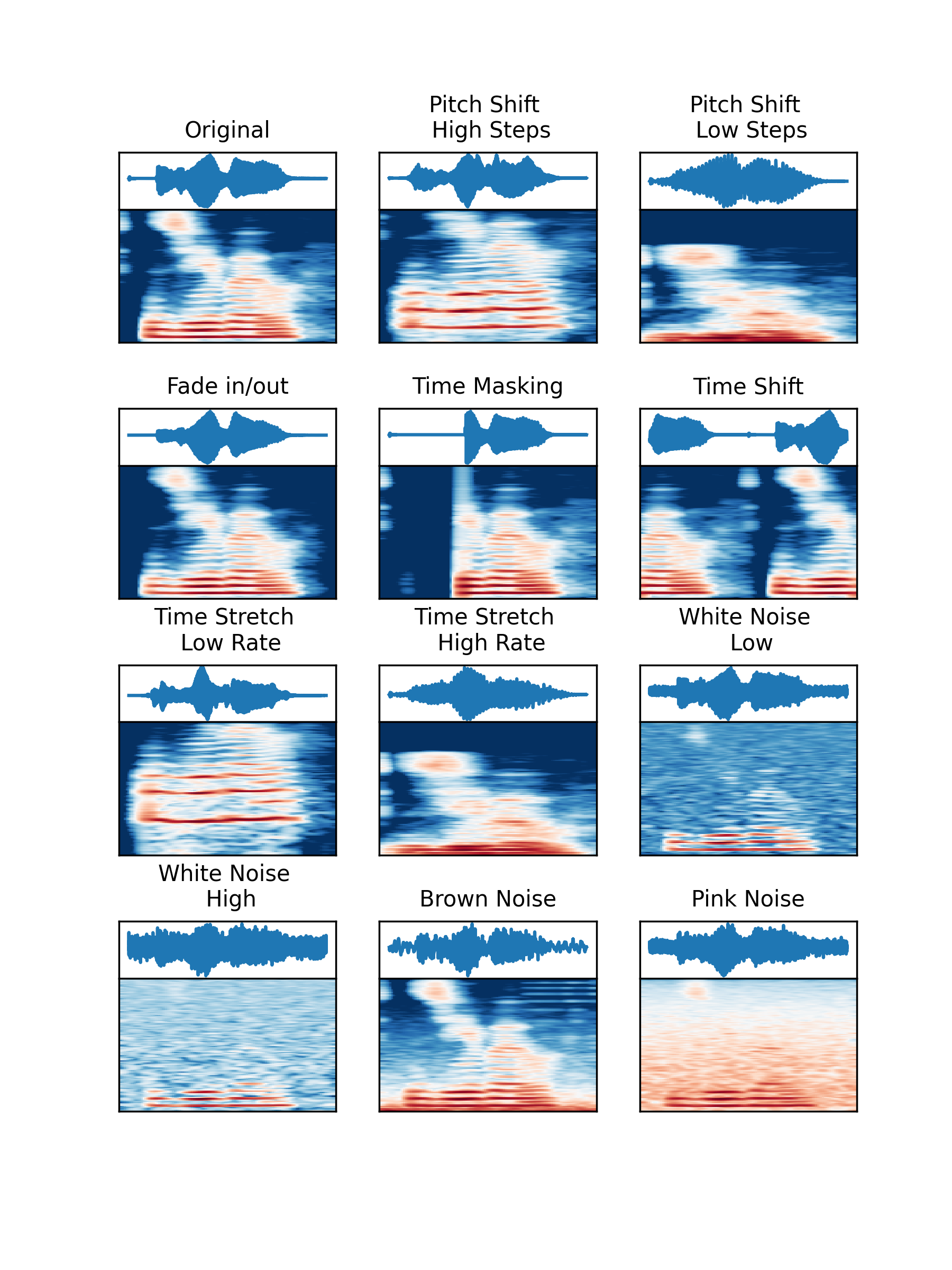}
\end{center}
\caption{Raw audio with the subsequent mel-spectrogram for each audio augmentation utilized in the paper. Each augmentation have a varying degree of the shown transformation depending on random hyper parameters. Here an example of each augmentation is illustrated. }\label{fig:augmentations}
\end{figure}

To investigate the impact of various data augmentations suitable for learning auditory representations, we deployed six distinct augmentations (Figure ~\ref{fig:augmentations}). Each augmentation was applied directly to the audio signal. Augmentations that directly influence spectrograms were not included \citep{Park_2019} to ensure that we could make direct comparison of augmentations performance both with the 1D and 2D models. Introduced augmentations could be categorised as either frequency or temporal transformations:

\begin{enumerate}
    \item Frequency Transformations
    \begin{enumerate}
        \item \textbf{Pitch Shift (PS)}: randomly raises or lowers the pitch of the audio signal \citep{brian_mcfee_2020_3955228}. Based on experimental observation, we found the range of pitch shifts that maintained the overall coherency of the input audio was in the range [-15, 15].
        \item \textbf{Fade in/out (FD)}: gradually increases/decreases the intensity of the audio in the beginning/end of the audio signal. The degree of the fade was either linear, logarithmic or exponential (applied with uniform probability of $1/3$). The size of the fade for either side of the audio signal could at maximum reach half of the audio signal. The size of the fade was another random parameter picked for each sample.
        \item \textbf{Noise Injection}: mix the audio signal with random white, brown and pink noise. In our implementation, the intensity of the noise signal was randomly selected based on the strength of signal-to-noise ratio. We adopted two versions of this augmentation: (1) applied only white noise with varying degree of intensity (White Noise), (2) applied either white, brown, or pink depending on an additional random parameter sampled from uniform distribution (Mixed Noise).
    \end{enumerate}
    \item Temporal Transformations
    \begin{enumerate}
        \item \textbf{Time Masking (TM)}: given an audio signal, in this transformation we randomly select a small segment of the full signal and set the signal values in that segment to normal noise or a constant value. In our implementation, we not only randomly selected the location of the masked segment but also we randomly selected the size of the segment. The size of the masked segment was set to maximally be $1/8$ of the input signal.
        \item \textbf{Time Shift (TS)}: randomly shifts the audio samples forwards or backwards. Samples that roll beyond the last position are re-introduced at the first position (rollover). The degree and direction of the shifts were randomly selected for each audio. The maximum degree that could be shifted was half of the audio signal, while, the minimum was when no shift applied to the signal.
        \item \textbf{Time Stretching (TST)}: slows down or speeds up the audio sample (while keeping the pitch unchanged). In this approach we transformed the signal by first computing the STFT of the signal, stretching it using a phase vocoder, and computing the inverse STFT to reconstruct the time domain signal \citep{brian_mcfee_2020_3955228}. Following those transformations, we down-sampled or cropped the signal to match the same number of samples as the input signal. When the $rate$ of stretching was greater than 1, the signal was sped up. Otherwise when the $rate$ of stretching was less than 1, then the signal was slowed down. The $rate$ of time stretching was randomized for each audio with range values of [0.5, 1.5].
    \end{enumerate}
\end{enumerate}

\begin{figure*}[h]
\begin{center}
\includegraphics[width=\linewidth]{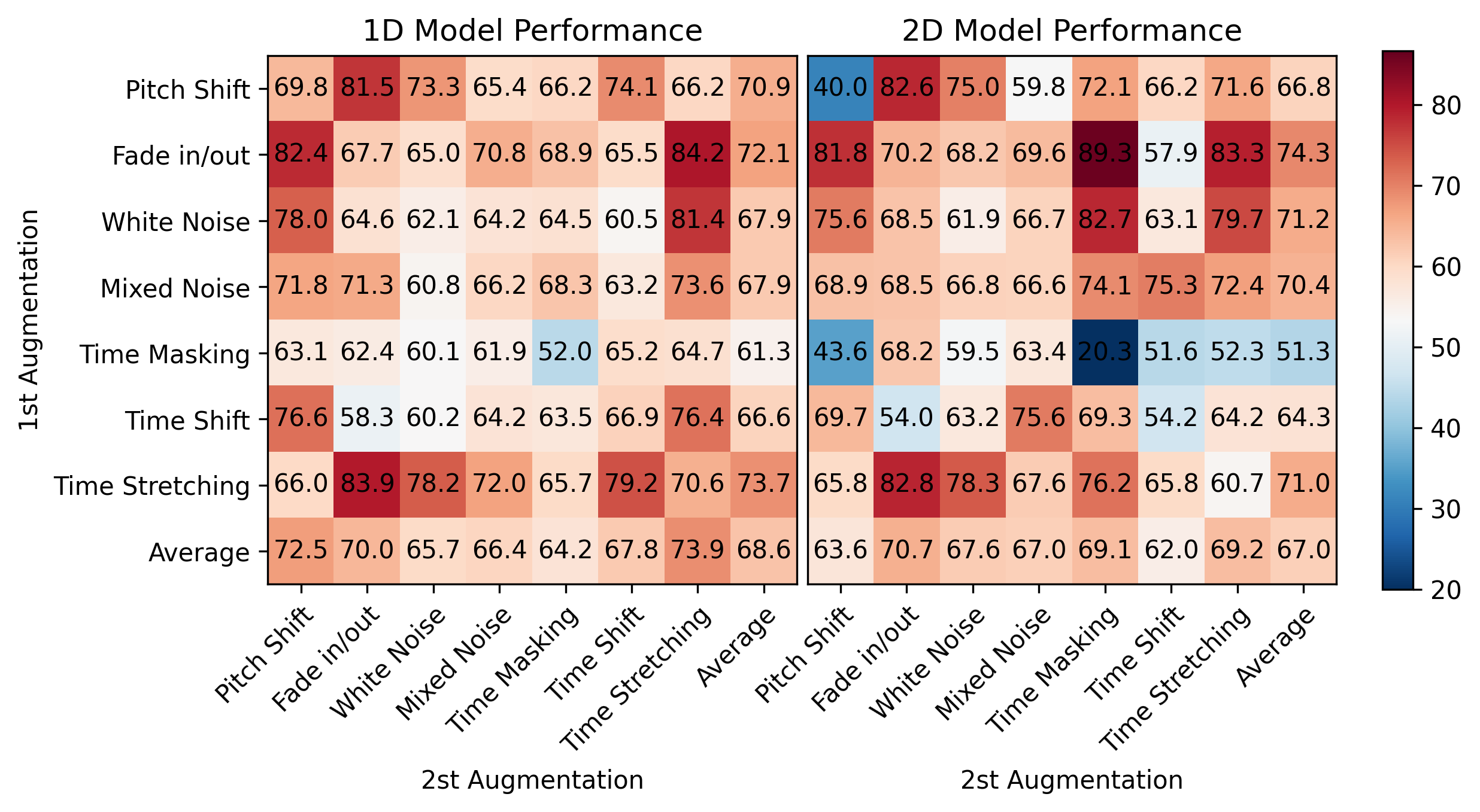}
\end{center}
\caption{Top-1 test performance on Speech Commands-10 for 1D (left matrix) and 2D (right matrix) models trained for 1000 epochs. The diagonal line represents the performance of single augmentation, while other entries represent the performance of paired augmentations. Each row shows the first augmentation and column shows the second augmentation applied sequentially.}\label{fig:augmentations_mat}
\end{figure*}
\subsection{Datasets} \label{subsec:datasets}

In this work, we evaluated the performance of our proposed framework on benchmark audio datasets in three different domains (speech, music, and environmental sounds). All datasets have predefined train-validation-test splits by authors (except the ECS dataset), we used the test splits to compute analysis. The datasets are:

\begin{itemize}
    \item \textbf{Speech Commands} (Speech): composed of 105,829 16kHz single-channel audios \citep{warden2018speech} from  2,618 speakers. Each audio file contains a one second recording of a single spoken English word from limited vocabulary. The dataset contains 35 labels (words) such as one-digit numbers, action oriented words, and arbitrarily short words. In addition to the full dataset, we derive a simpler version ($\sim$20k samples) with only the utterances of the one-digit numbers.
    \item \textbf{NSynth} (Music): contains 305,979 four seconds audio of musical notes, each with a unique pitch and musical instrument family \citep{nsynth2017}. For every musical note, the note was held for the first three seconds and allowed to decay for the final second. Similar to Speech Commands, we composed two variations of the same dataset, (1) we utilized musical instrument family as the class labels (11 classes) and (2) we used pitch as the class labels (128 classes). Both variations of NSynth dataset included the same amount of data, however, the number of classes varied. 
    \item \textbf{Dataset for Environmental Sound Classification} (Environmental): consists of two variants ESC-10 and ESC-50 provided by the authors. Similar to previous datasets described, the two variants describe the number of classes. The ESC-50 dataset consists of 2,000 5-seconds environmental recordings equally distributed across 50 classes (40 clips per class). Classes such as animal, natural and water, non-speech human, interior and exterior sounds \citep{YDEPUT_2015}. The ESC-10 is a subset of ESC-50 consisting of 400 recordings, making it the dataset with the least number of training data in our collection. Both datasets are divided into 5 folds by the authors. In this work, we utilized the first 4 folds for training and the last for testing.
\end{itemize}

\section{DATA AUGMENTATIONS FOR CONTRASTIVE LEARNING}\label{sec2}

Data augmentations have been widely adopted in audio \citep{mcfee2015software, schluter2015exploring, ko2015audio, Park_2019} and image \citep{krizhevsky2012imagenet, henaff2019data} domains. Moreover, recently it has been shown that some sequences of augmentations in image domain can offer a relatively better performance compared to other augmentations in contrastive learning \citep{chen2020simple}. In this section, we investigate the impact of different augmentations applied to the signal level on the quality of learned auditory representations. 

To investigate the effect of individual auditory data augmentations and their sequential ordering, we perform comprehensive training of SimCLR framework on Speech Commands-10 dataset for 1000 epochs on all the proposed auditory augmentations (Section \ref{sec:aug}). Figure \ref{fig:augmentations_mat} shows top-1 test performance on both 1D and 2D variants of ResNet18. The diagonal line represents the performance of single augmentation, while other entries represent the performance of paired augmentations.  Each row indicates the first augmentation and each column shows the second augmentation applied sequentially. The last column and row in each matrix represents the averaged predictive performance of a specific augmentation. The last column depicts the average when the augmentation was applied first, while the last row shows the average when the corresponding augmentation was applied second. The bottom right element represents the average of the whole matrix. Similar to \cite{chen2020simple}, we found that multiple augmentations are required to learn efficient representations. In particular, with the 1D model, we observe that the composition of fade in/out, time stretching and pitch shifting significantly are the top-3 augmentations that improve the quality of representations. While on the 2D model, we observe that fade in/out, time masking and time shifting are premier in improving the quality of the auditory representations. On average the 1D variant of the model shows better performance (\textbf{1D}: $68.6 \pm 0.82$; \textbf{2D}: $67.0 \pm 1.36$). However, the 2D variant of the model achieves the maximum recorded performance (89.3\%).


\section{RAW SIGNAL VERSUS TIME-FREQUENCY FEATURES}\label{sec3}

Time-frequency audio features have been used extensively in the literature, as algorithms trained on such features have consistently demonstrated better performance compared to algorithms trained on raw audio signals \citep{tokozume2017learning, tokozume2017learninga, Guzhov2020ESResNetES, nsynth2017}. In this section, we extend on this concept by investigating the efficiency of representations learned from raw signal and time-frequency features.

Table \ref{tab:performance1v2d} shows the accuracy of evaluation head attached to the frozen encoder trained with augmentations that yielded the best performance from section \ref{sec2}. We found that time-frequency features compared to raw audio signal consistently improve the learned representations. In particular, fade in/out and time masking using time-frequency features outperforms all other methods. We also found that increasing the number of augmentations does not necessary improve the learned representations. This was observed when predictive performance degraded after appending time-stretching augmentation to fade in/out and time masking.
\begin{table*}[t]
\caption{Accuracy of ResNet18 trained on various datasets using different augmentations. }
\label{tab:performance1v2d}
\begin{center}
\begin{tabular}{lcccccc}
\hline
\multicolumn{1}{c}{Models} & \multicolumn{6}{c}{Datasets}                                                                                                   \\ \hline
\multicolumn{1}{c}{1D}     & ECS-10        & ECS-50        & SC-10         & SC-50         & \multicolumn{1}{l}{NSynth-11} & \multicolumn{1}{l}{NSynth-128} \\
FD                         & 12.5          & 3.3           & 67.7          & 16.1          & 25.1                          & 52.2                           \\
FD + TST                   & 52.1          & 2.9           & 84.2          & 24.2          & 34.6                          & 70.3                           \\
FD + TST + PS              & 10.0          & 3.5           & 79.9          & 23.6          & 46.2                          & 10.8                           \\ \hline
\multicolumn{1}{c}{2D}     &               &               &               &               &                               &                                \\
FD                         & 46.2          & 19.5          & 70.2          & 13.7          & 31.1                          & 52.8                           \\
FD + TM                    & \textbf{68.7} & 33.7          & \textbf{89.3} & \textbf{29.1} & \textbf{48.8}                 & \textbf{84.1}                  \\
FD + TM + TS               & 62.5          & \textbf{40.4} & 84.6          & 25.0          & 48.6                          & 80.4                           \\ \hline
\end{tabular}
\end{center}
\end{table*}

{
\renewcommand{\arraystretch}{1.5}
\begin{table}[t]
\caption{Accuracy of ResNet18 trained for 1000 empochs on Speech Command-10 dataset with incrementally less labels. During evaluation phase, we trained the evaluation head on all the labeled data with the frozen encoder.}
\label{tab:performancess}
\begin{center}
\begin{tabular}{lcccl}
\hline
\multicolumn{1}{c}{Training Method} & \multicolumn{4}{c}{Labeled Data Percentage}          \\ \hline
\multicolumn{1}{c}{}                & 100\%         & 20\%          & 10\%          & 1\%  \\ \cline{2-5} 
Supervised                          & 94.9          & 86.4          & 68.4          & 28.6 \\ \cline{2-5} 
Self-Supervised                        & \multicolumn{4}{c}{84.8}                             \\ \cline{2-5} 
Semi-Supervised                    & \textbf{96.1} & \textbf{90.5} & \textbf{89.1} & 77.9 \\ \hline
\end{tabular}
\end{center}
\end{table}
}

\begin{figure}[h]
\begin{center}
\includegraphics[width=\linewidth]{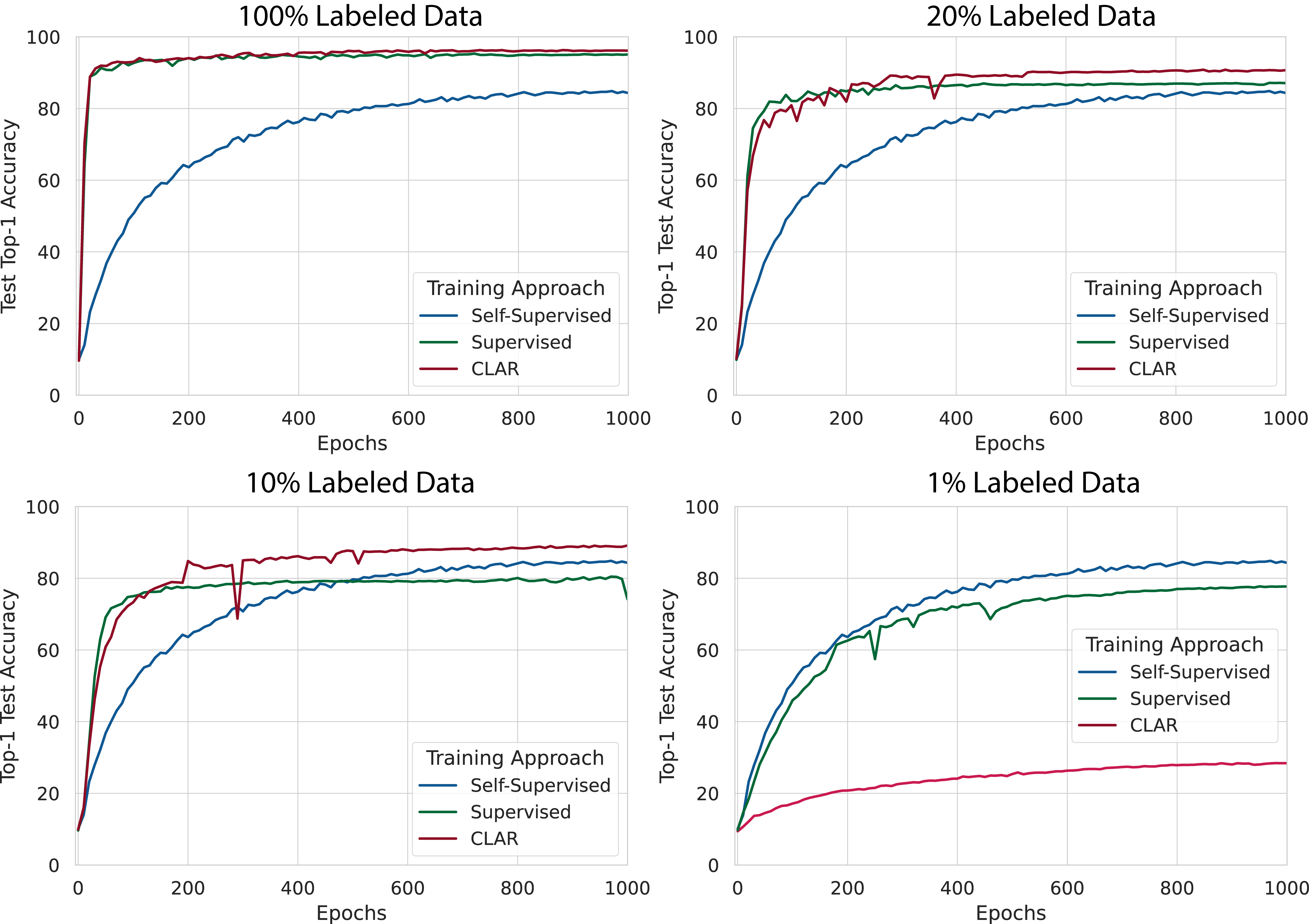}
\end{center}
\caption{Top-1 test performance on Speech Commands-10 computed every 10-epochs by training an evaluation head attached to a frozen encoder over 1000 epochs. Each sub-figure represents the top-1 test performance on varying percentage of labeled data.}\label{fig:speed}
\end{figure}
\section{CLAR REPRESENTATIONS VERSUS SUPERVISED AND SELF-SUPERVISED}\label{sec4}

In this section, we utilize augmentations that yielded the best performance on SC-10 from section \ref{sec3} to investigate the efficiency of CLAR with supervised and self-supervised methods. We utilize SC-10 as the dataset of choice and investigate both the predictive performance of the evaluation head (Section \ref{sec:train} for more explanation) every 10 epochs for a maximum of 1000 epochs. This would not only shed light on the final performance but also the speed at which the methods reach such performance. Moreover, as CLAR is capable of semi-supervised training, we test its capability by training models on 100\%, 20\%, 10\% and 1\% labeled data. For our self-supervised method we use no labeled data, hence, the performance would be the same across the board.

\subsection{CLAR improves learned representations}

Table \ref{tab:performancess} shows the top-1 accuracy of models trained using various methods while changing the percentage of labeled data. We found that the representations of the encoder trained with the CLAR method outperforms the representations learned using supervised and self-supervised (SimCLR) methods when trained over the same number of epochs. The efficiency of the representations is demonstrated when all the methods are trained on 100\% of the labeled data (96.1\%). Furthermore, we show that this trend continues as we decrease the labeled data. That is, while the supervised methods loses ~66\% of the performance as we decrease the labeled data from 100\% to 1\%, CLAR decreases by only ~19\%. These results show that CLAR indeed combines both supervised and self-supervised methods to draw more efficient representations. Lastly, we found that when we decrease the amount of labeled data to only 1\%, the performance of CLAR's learned representation degrades compared to the self-supervised method. This could be the result of overfitting more to the labeled data resulting in less efficient representations.

\subsection{CLAR improves the speed of learning representations}

Figure \ref{fig:speed} shows the top-1 test performance of SC-10 dataset computed every 10-epochs by training an evaluation head attached to a frozen encoder over 1000 epochs. We found that the CLAR method not only improves the representations in the encoder but also improves the speed at which those representation are learned compared to the self-supervised approach. In particular, when we provide 100\% of the labeled data, CLAR shows not only better predictive performance compared to supervised and self-supervised method but also show better training speed than self-supervised. As we decrease the amount of labeled data, this trend continues while the gap between the supervised method and CLAR test performance increases. These results suggest that both the self-supervised and supervised training tasks indeed share common representations that improves latent representations in the encoder. Moreover, the training algorithm optimize for the CE loss first followed by gradual slow optimization of the NT-Xent loss.

\section{CONCLUSION}

In this paper, we demonstrated the success of contrastive learning in earning efficient auditory representations. Our extensive and comprehensive experiments on various design choices revealed the effectiveness of our proposed framework (CLAR) in terms of recognition performance as well as reduced training time in comparison with supervised and self-supervised methods. Together, our results depicts a promising path towards automated audio understanding.

\bibliography{references}
\end{document}


%

%

\onecolumn
\aistatstitle{Supplementary Materials}
Codebase for CLAR a contrastive learning framework for auditory representations. In the code provided, we only have PyTorch implementation. Please follow the steps below to ensure that the code work correctly.

\section{Install Python requirements}

Before running the scripts, install all the python libraries:

\begin{lstlisting}
pip install -r requirements.txt
\end{lstlisting}
 
Additionally, we pre-installed cuda/10.1, cudnn/7.6.5, and nccl/2.5.6. Instructions can be found at \url{https://developer.nvidia.com/cuda-toolkit-archive}

\section{Datasets}

Most adopted dataset in the paper is \href{http://deepyeti.ucsd.edu/cdonahue/wavegan/data/sc09.tar.gz}{Speech Commands Zero through Nine (SC09)}. This is a subset of the \href{https://ai.googleblog.com/2017/08/launching-speech-commands-dataset.html}{Speech Commands Dataset}. Before we utilize the dataset, we convert it to `lmdb` format for fast loading. If you would like to convert your own dataset to `lmdb` format, you could use `DatasetConverter.py`. Alternatively, you can adjust the method `load\_datasets` in the code to use any dataset using the PyTorch framework.

We have provided the test data in `lmdb` format for SC09 for testing purposes. You can download it from \url{https://drive.google.com/file/d/17TQvtKf1M3Mbm46uzd9SfXVBlrdB-Y9l/view}

Once you have the lmdb files, create a directory named `data` and put those files in there.

\section{Training}

There are 3 scripts for different methodology used in the paper: `main\_supervised.py`, `main\_unsupervised.py` and `semi\_supervised.py`

To start training with default options, this command will do:

\begin{lstlisting}
python main_supervised.py
\end{lstlisting}

The command above will train a 1D version of ResNet18. To change that, you can replace the name:

\begin{lstlisting}
python main_supervised.py --model_name='2d'
\end{lstlisting}

As long as we have `2d` in the name, the model will use the 2D version of the model. If you would like to change the dataset used, you will need to replace the name:

\begin{lstlisting}
python main_supervised.py --model_name='2d' --dataset='sc09'
\end{lstlisting}

This expect that you have `train\_sc09.lmdb`, `valid\_sc09.lmdb` and `test\_sc09.lmdb` in your data folder.

To add augmentations to the training, you will need to pass parameters that reflect a given augmentation. The set of augmentations with the best results are as follows:

\begin{lstlisting}
python main_supervised.py --model_name='2d' --dataset='sc09' --tm=True --fd=True
\end{lstlisting}

\section{Testing}

To get the test accuracy, all you need is to set the `train` parameter to `False`:

\begin{lstlisting}
python main_supervised.py --model_name='2d' --dataset='sc09' \
--tm=True --fd=True --train=False
\end{lstlisting}

Make sure that you have the saved model in the `./model` directory. 

Please note that depending on the augmentations and dataset used the model name will change. For instance, for the above command the model name will be `2d\_100\_fd\_tm`. This was done to save models for different augmentations and datasets without losing track of results. Hence, the script will look for the saved model at `./models/2d\_100\_fd\_tm/sc09/`. The `100` in the number indicate that we used 100\% of the labeled data to train the model.

\section{Saved Models}

Models for the final results can be downloaded from \url{https://drive.google.com/drive/folders/1GA-HCBXlRuOCWHQ-_A_Ei265OEek2kms?usp=sharing}

Please make sure to maintain the folder structure.